# Machine learning and evolutionary algorithm studies of graphene metamaterials for optimized plasmon-induced transparency

Tian Zhang[1] ·Qi Liu[1] ·Yihang Dan[1] ·Shuai Yu[1] ·Xu Han[2] ·Jian Dai[1] ·Kun Xu[1]

[1] State Key Laboratory of Information Photonics and Optical Communications, Beijing University of Posts and Telecommunications, Beijing 100876, P.R. China
[2] Huawei Technologies Co., Ltd, Shenzhen 518129, Guangdong, P.R. China

E-mail: xukun@bupt.edu.cn

**Abstract**
Machine learning and optimization algorithms have been widely applied in the design and optimization for photonic devices. In this article, we briefly review recent progress of this field of research and show some data-driven applications (e.g. spectrum prediction, inverse design and performance optimization) for novel graphene metamaterials (GMs). The structure of the GMs is well-designed to achieve the wideband plasmon induced transparency effect, which is regarded as optimization object and can be theoretically demonstrated by using transfer matrix method. Some classical machine learning algorithms, including *k* nearest neighbour, decision tree, random forest and artificial neural networks, are utilized to equivalently substitute the numerical simulation in the forward spectrum prediction and complete the inverse design for the GMs. The calculated results demonstrate that all the algorithms are effective and the random forest has advantages in terms of accuracy and training speed. Moreover, the single-objective and multi-objective optimization algorithms are used to achieve steep transmission characteristics by synthetically taking many performance metrics into consideration. The maximum difference between the transmission peaks and dips in the optimized transmission spectrum can reach 0.97. In comparison to previous works, we provide a guidance for intelligent design of photonic devices and advanced materials based on machine learning and evolutionary algorithms.

(Some figures may appear in color only in the online journal)

## 1. Introduction

Traditionally, the design and optimization of photonic devices depends on the repeated trial or physics-inspired methods [1-2]. With the increase of performance metric and integration level, the design and optimization processes for photonic devices become computationally expensive and complex [3]. For example, owning to the excellent electronic and optical properties [4-9], graphene, a typical 2D materials [10], has been applied in many photonic devices, such as optical modulator [11], photoelectric detector [12], sensor [13], absorber [14], switching [15], polarization controller [16], diode [17] and so on. For a graphene nanostructure, we usually consider the influence of critical physical parameters (e.g. the chemical potential and the number of layers) on the electromagnetic responses. Nevertheless, the lack of the empirical relationships between physical parameters and corresponding electromagnetic responses often leads to the time-consuming brute force search, which calculates the electromagnetic responses for all physical parameters by using the numerical simulations, such as finite-difference time-domain (FDTD) and finite element method (FEM) [18-19]. In fact, we can construct a theoretical model to describe the physical mechanism behind the physical phenomenon [20]. The electromagnetic responses for different physical parameters can be quickly calculated based on the theoretical model. However, it should be noted that the construction of the theoretical models for complex graphene nanostructures is generally difficult because the physical mechanisms are hard to understand. In order to solve the above problems, some data-driven approaches based on machine learning technique have been proposed to equivalently substitute the numerical simulation or even theoretical model. Especially in recent years, with the development of high performance computing, artificial neural networks (ANNs), deep learning in particular, have attracted a great deal of research attentions for an impressively large number of applications, including image processing [21], natural language processing [22], acoustical signal processing [23], time series processing [24], self-driving [25], game [26], robot [27] and so on. Many researchers attempt to use the ANNs to construct a model describing the relationship between the physical parameters of photonic devices and electromagnetic responses [28-50]. Once the data-driven model is constructed, when the physical parameters are inputted into the model, the electromagnetic responses can be calculated in a very short time based on the model inference [20, 28], Typically, the inference time of the model is significantly smaller than the calculation time of the numerical simulation. Thus, the equivalent approximation of the numerical simulation based on a black-box model can accelerate the processes of device-level variability analysis and performance evaluation for photonic devices. In addition, it has been demonstrated that the data-driven methods are conductive to the inverse design of photonic devices. The purpose of the inverse design is to search for the suitable physical parameters, which can generate the targeted electromagnetic response [1]. Generally speaking, the inverse design of photonic devices are solved by using optimization algorithms. And if the the potential relationship between the electromagnetic responses and physical parameters can be constructed by using machine learning technique, the inverse design problems are also solved by the data-driven methods. Contrary to the model used in the simulation approximation (it predicts the electromagnetic responses from the physical parameters), the model used in inverse design predicts the physical parameters according to electromagnetic response. We briefly review recent progress of this field of research.

For example, J. Peurifoy *et al.* found that the ANNs could be used to simulate the light scattering and inversely determine the physical parameters of multilayer nanosphere [28]. Here, the electromagnetic responses for all physical parameters of nanospheres were predicted by the ANNs, which were trained by using a small sampling of simulation results. And they pointed out that the ANNs had ultra-fast prediction speed in comparison to numerical simulation. It should be noted that the principles behind the machine learning techniques that were used in the approximation of numerical simulation and the inverse design of photonic devices were supervised learning or data regression between physical parameters and electromagnetic responses. Many researcher begun to explore the applications of supervised learning or data regression in the design of photonic devices from two perspectives: photonic devices and algorithms. For the aspect of photonic devices, the shallow ANNs were also used in the design and optimization for plasmonic waveguide system [20], metal grating [29], $VO_2$-based nanostructure [30], strip waveguide [31], chirped Bragg grating [31], sub-wavelength grating coupler [32], plasmonic nanoparticle [33] and so on. In the above works, the successful applications of the ANNs demonstrated that the relatively simple network architectures were enough to fit a small quantity of physical parameters. For the aspect of algorithms, the ANNs were meticulously designed to fit various photonic devices. Many ANNs with different network architectures, for example deep neural networks [34-41], adaptive neural networks [42], bidirectional neural networks [43-44], tandem networks [45] had been proposed to design and optimize for the complex photonics devices and optical properties, including power splitters [34], metasurfaces [35-36], plasmonic colours [37], photonic crystal nanocavities [38-39], optical chirality [40], plasmonic sensor [41], metamaterials [42-43], silicon color [44], multilayer film [45] and so on. There is no doubt that for the complex photonic devices, such as a silicon resonator based on encoding metamaterials with random distribution [34], deep neural networks (deep learning) were an effective modelling method to construct the complex relationship between physical parameters and electromagnetic responses. In order to reduce the size of training set and improve the accuracy for the ANNs, Y. Qu *et al.* proposed to use transfer learning technique to migrate knowledge between different physical scenarios in inverse design [46]. More interestingly, as a unsupervised learning method, the generative adversarial networks (GAN) in deep learning had been proven effective in the generation of high performance photonic devices with a broad design space [47-50]. Recently, J. Jiang *et al.* found that a high efficiency, topologically complex device could be produced by the GAN with a wide parameter space [50]. In addition, except for the supervised learning and unsupervised learning, reinforcement learning was also used to build an autonomous system to solve the decision-making problems in the optimization of photonic devices [51-55]. Recently, I. Sajedian *et al.* used the deep reinforcement learning to search for the optimal material types and structural parameters of the high-quality metasurface [54]. In order to speed up the search process of training sets and optimize for the neural network architectures, the ANNs were combined with evolution algorithms to design the micro-to-nano photonic couplers [56, 57]. Besides, other machine learning techniques, including dimension reduction and bayesian optimization, were also used to design grating coupler and wavelength-selective thermal radiator [58, 59]. It should be noted that although the ANNs provided an effective approximation approach to replace the numerical simulation, it required a great deal of time to collect training sets. And compared with classical machine learning algorithms, such as support vector machines (SVM) and random forest (RF), the ANNs had disadvantages in training time. It had been demonstrated that classical machine learning algorithms were more effective in some uncomplicated applications with a small quantity of physical parameters [60, 61]. However, there was a lack of comprehensive analytical report for the applications of machine learning algorithms in the simulation approximation and inverse design for photonic devices.

In addition to the data-driven methods mentioned above, the inverse design of photonic devices could be solved by using optimization algorithms, which were divided into two classes: gradient based methods and gradient free methods [20]. As a representative method of gradient based methods, adjoint variable method (AVM) could not only design for the linear optical devices but also optimize for the nonlinear devices in the frequency domain [62-63]. In 2018, T.W. Hughes *et al.* proposed a novel training method to compute the gradients of weights for optical neural networks based on AVM [64]. The objective-first optimization method and the steepest descent method were used to design for wavelength demultiplexer and computational metasturctures [65-66]. However, gradient based methods, such as AVM, required physical background to derive the gradient of objective function. Obviously, it was relatively complex and increased the difficulty in the practice application of gradient based methods. On the other hand, as the representative algorithms of gradient free methods, evolution algorithms (e.g. genetic algorithms) and search algorithms (e.g. direct-binary search

[67]) were used in the inverse design and optimization for photonic devices [67-71]. Recently, we used two kinds of evolution algorithms, genetic algorithms (GA) and particle swarm optimization, to determine the hyper-parameters and weights of the optical neural networks [70]. A segmented hierarchical evolutionary algorithm was established to solve the large-pixelated and complex inverse meta-optics design [69]. Notably, although evolution algorithms had advantages in simplicity and effectiveness, they easily fell into local optimum and demanded significant computing time [72]. Quantum genetic algorithm (QGA), which took advantage of the power of quantum computation, had been demonstrated to speed up genetic procedures. And no matter gradient based methods and gradient free methods, they usually optimized for a single performance metric of photonic devices, such as coupling efficiency. However, multi-objective optimization algorithms, which could synthetically optimize for multiple performance metrics, were rarely used to find the suitable physical parameters of photonic devices.

In this article, we provide a guidance for the intelligent design of photonic device based on machine learning and evolution algorithms. Recently, various structures based on graphene and surface plasmon polaritons (SPPs) [73-75], for example graphene metamaterials (GMs) [76-79], graphene nanoribbons (GNRs) [80-82] and graphene waveguide [83-85], have been proposed to construct plasmonic filter [83], perfect absorber [80], sensor [81], logic gate [85] and so on. In these structures, the GMs that consists of periodically spaced GNRs have attracted widespread attention because of their relatively simple fabrication technique [80-82]. In order to evidently show the design effects, we propose novel GMs consisted of parallel GNRs and try to obtain spectrum prediction, inverse design and performance optimization for the GMs. In addition, the physical parameters of the GMs are well-designed to achieve the plasmon induced transparency (PIT) effect in the transmission spectrum. The reason for the selection of the PIT effect as optimization object is attributed to that it can construct high-performance photonic devices, such as plasmonic filter, switching, polarization-insensitive sensor (parallel GNRs) and perfect absorber (crossed GNRs) [18, 19, 86]. Several classical machine learning algorithms, including $k$ nearest neighbour (kNN), decision tree (DT), extremely randomized trees (ERT), random forest (RF), and ANNs are used to achieve forward spectrum prediction and inverse design for the GMs. The calculated results exhibit that all the regression algorithms are effective and the RF has advantages in terms of accuracy and training speed. On the other hand, although the single-objective optimization have been used to optimize for wavelength multiplexer [87], mode multiplexer [88], polarization beam splitter [67], polarization rotator [68], power splitter [69], previous researches pay little attention to the design of graphene nanostructures, especially for the optimization of graphene. We use the single-objective optimization and multi-objective optimization algorithms to optimize for the GMs by synthetically taking many different performance metrics into consideration. The maximum difference between the transmission peaks and dips in the optimized transmission spectrum can reach 0.97.

## 2. Device design and simulation results

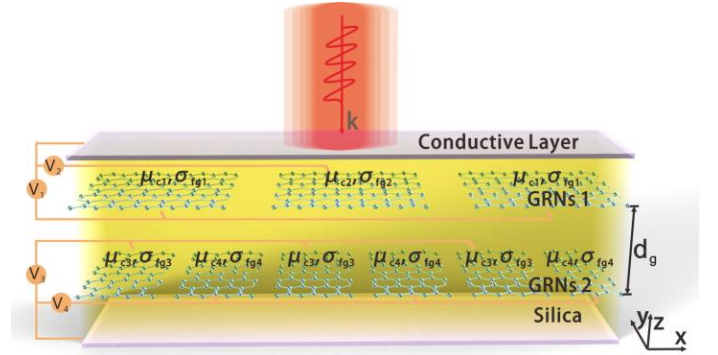

**Figure 1.** The schematic view of the proposed GMs, which consist of double layer GNRs embedded into insulated silica layer with a separation $d_g$=300 nm.

As shown in Figure 1, our proposed novel GMs consist of two layer GNRs with alternative chemical potentials. Here, the double layer GNRs in the GMs are periodically arranged and infinite along x (y) axis. The thin conductive layer covered on the bottom and top of silica layer ($n_{sio2}$=1.45) forms as electrodes to alternatively apply voltage $V_1$ ($V_3$) and $V_2$ ($V_4$) on the GNRs 1 (2), leading to the graphene ribbons of two GNRs with alternative chemical potential ($\mu_{c1}$ and $\mu_{c2}$ for GNRs 1, $\mu_{c3}$ and $\mu_{c4}$ for GNRs 2). The period of the GNRs 1 (GNRs 2) is set as $\Lambda_1$=400 nm ($\Lambda_2$=200 nm), and the width of the graphene ribbon in GNRs 1 (GNRs 2) is $w_1$=350 nm ($w_2$=175 nm), leading to a filling ratio of $r_1$=0.875 ($r_2$=0.875). We employ the Kubo formula to model the conductivity of the GNRs in the numerical simulation of the GMs [75]:

$$\sigma_g = i\frac{e^2 k_B T}{\pi \hbar^2 (\omega + i\tau^{-1})}\left[\frac{\mu_c}{k_B T} + 2\ln\left(\exp\left(-\frac{\mu_c}{k_B T}\right)+1\right)\right] + i\frac{e^2}{4\pi\hbar}\ln\left[\frac{2|\mu_c| - \hbar(\omega + i\tau^{-1})}{2|\mu_c| + \hbar(\omega + i\tau^{-1})}\right] \quad (1)$$

where $k_B$, $T$ (=300 K), $\hbar$, $\tau$ (=0.5 ps), $\mu_c$, $e$, and $\omega$ represent the Boltzmann's constant, temperature, reduced Planck's constant, relaxation time, chemical potential, electron charge and angular frequency, respectively. For a few layers (<6) of graphene, the conductivity of it can be expressed as $\sigma_{fg}=N\sigma_g$, where $N$ is the number of layers [16]. In the mid-infrared, the simplified conductivity can be calculated by considering the domination of interband electron-photon process and $\mu_c \gg k_B T$

$$\sigma_{fg} = i\frac{Ne^2 \mu_c}{\pi \hbar^2 (\omega + i\tau^{-1})} \quad (2)$$

In order to analyze the excitation condition of the SPPs in the GMs, the dispersion equation is retrieved based on the Maxwell equation and continuous boundary condition [75]

$$\frac{\varepsilon_1}{\sqrt{\beta_{SPP}^2 - \frac{\varepsilon_1 \omega^2}{c^2}}} + \frac{\varepsilon_2}{\sqrt{\beta_{SPP}^2 - \frac{\varepsilon_2 \omega^2}{c^2}}} = -\frac{i\sigma_{fg}}{\omega \varepsilon_0} \quad (3)$$

where $\beta_{SPP}$ is the propagation constant of SPPs, $c$ represents the light speed in vacuum, $\varepsilon_0$ is the dielectric constant of free space, $\varepsilon_1$ and $\varepsilon_2$ are the effective permittivities of the medium on each side of GNRs ($\varepsilon_1$ and $\varepsilon_2$ are equal to $\varepsilon_{SiO2}=n_{SiO2}^2=2.1$ because the GNRs are surrounded by the same medium in our proposed GMs). Here, since the solution satisfies $\beta_{SPP} \gg \omega/c$, the effective refractive index of SPPs deduced from Eq. (3) is given by [74]

$$n_{SPP} = \frac{\beta_{SPP}}{k_0} = \frac{2\varepsilon_0 \varepsilon_{SiO2}\pi\hbar^2 c}{Ne^2\mu_c}(\omega + i\tau^{-1}) \quad (4)$$

where $k_0$ relates to the wave vector of vacuum. As shown in Figure 2(a)-(b), the dispersion curves (solid lines) of few graphene layers surrounded by silica match well with the dispersion data (circle marks) calculated by the mode solution. Notably, the monolayer and multilayer graphene in the numerical simulation are treated as a surface with electric conductivity $\sigma_{fg}$ since the graphene ribbons are ultrathin. It can be found that both the real part and imaginary part of the effective refractive indices for the SPPs on graphene layers decrease with the increasing of wavelength. Thus, when the wavelength of incident light increases, the field confinement (propagation loss) of the SPPs on graphene becomes weaker (smaller). In addition, the field confinement of the SPPs on graphene becomes weaker with the increasing of the number of graphene layers. In order to increase the interaction of the upper GNRs and the lower GNRs, we set the number of the GNRs as $N$=4 in the following article.

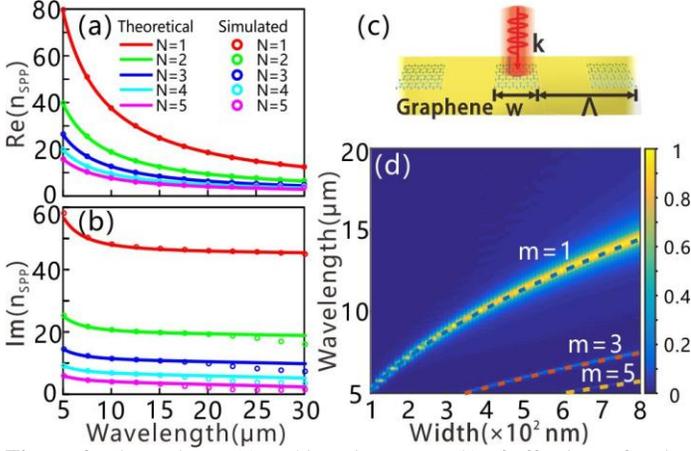

**Figure 2**. The real part (a) and imaginary part (b) of effective refractive indices for SPPs. The solid lines are the dispersions of SPPs calculated by theoretical model, and the marks are those calculated by the mode solution. (c) Schematic of the single layer grating consisted of GNRs. (d) When mid-infrared wave normally incident on the single layer grating, blue, red and orange lines are the resonance curves for three modes $m$=1, 3 and 5, respectively. For comparison, the absorption contour patterns of the single layer grating are calculated by the FDTD method. The value of $w/\Lambda$ is set as 1/4. In (a), (b) and (d), the chemical potentials of graphene $\mu_c$ are set as 0.5 eV.

First of all, as shown in Figure 2(c), we analyze the transmission spectrum of a single layer grating composed of GNRs. For the grating with small filling ratio ($w/\Lambda$), the SPPs on a graphene ribbon can hardly interact with that on the adjacent graphene ribbon. Thus, the propagation of SPPs on the grating can be equivalently substituted by that in a single graphene ribbon. It has been demonstrated that the SPPs on a single graphene ribbon are nearly totally reflected at the boundary together with a phase jump of $\varphi$=0.27$\pi$ [89]. Thus, the SPPs excited on a graphene ribbon are caused by the Fabry-Perot (FP) like resonances, which needs to satisfy

$$\text{Re}(n_{SPP})k_0 w + \varphi = m\pi, \quad m=1,2,3,... \quad (5)$$

Substituting Eq. (4) into Eq. (5), the resonance frequency of the SPPs on GNRs can be achieved as following

$$\omega_r = \sqrt{\frac{(m-\varphi)Ne^2\mu_c}{2\varepsilon_0\varepsilon_{SiO2}\hbar^2 w}}, \quad m=1,2,3,... \quad (6)$$

As shown in Figure 2(d), the resonance curves (three dashed lines) for three modes of the single layer graphene grating agree with the numerical simulation results (the absorption contour patterns). Obviously, the comparison results verify the effectiveness of theoretical model (FP) and numerical simulation (FDTD). Here, we only calculate the resonance curves for the odd modes in Figure 2(d) since the even modes cannot be excited with normal incident wave [89].

To analyze the mechanism of the PIT effect that would emerge in transmission spectrum, the optical responses of the GMs which only contain the upper GNRs or the lower GNRs are calculated by using the FDTD method, respectively. The 2D simulations whose $x$ direction is set as periodic boundary conditions and other boundaries are set as perfectly matched layer are employed to simulate the GMs. The Fermi levels of GNRs are set as $\mu_{c1}$=0.7 eV, $\mu_{c2}$=0.5 eV, $\mu_{c3}$= 0.15 eV and $\mu_{c4}$= 0.75 eV in our simulation. As shown in Figure 3, when the TM polarized light normally illuminates on the GMs that only includes the upper GRNs (blue dashed line) and the lower GRNs (green dashed line), two obvious dips emerge in the transmission spectrum. Here, the appearances of the dips are related to the excitation of the SPPs modes on the GNRs. Next, we proceed to consider the optical characteristics of the complete GMs that includes the upper GNRs and the lower GNRs. From Figure 3, two transmission peaks respectively located between two dips emerge in the transmission spectrum, indicating the appearance of the PIT-like effect [7]. The PIT-like effect can be applied in the optical switching and slow light because it has large extinction ratio and wide bandwidth [90]. The optical characteristics of the dips in the PIT-like effect are similar to those of the single layer GNRs mentioned, which suggests that the appearance of dips are attributed to the excitation of the SPPs mode on GNRs. And the formation of the PIT-like effect can be explained by the normalized magnetic field distribution of transmission peaks (B and E) and dips (A, C, D and F). As shown in Figure 3, it can be observed that the appearances of dips A, C, D and F are related to the excitation of SPPs on the graphene ribbons 1, 2, 3 and 4, respectively. While it's the coupling between the SPPs mode on graphene ribbons 1 (2) and 3 (4) gives rise to transmission peaks B (D). In order to model the dynamic transmission of the GMs, the transfer matrix method is used to explain the physical phenomenon, The transfer matrix can be defined as [90]

$$H = M_2 S_{12} M_1 \quad (7)$$

where $M_1$ ($M_2$) and $S_{12}$ represent the matrices of the upper (lower) GNRs and silica, respectively. They are governed by

$$S_{12}=\begin{pmatrix} e^{i\varphi'} & 0 \\ 0 & e^{-i\varphi'} \end{pmatrix}, \quad M_q = \frac{1}{t_{21}}\begin{pmatrix} t_{12}t_{21}-r_{12}r_{21} & r_{21} \\ -r_{12} & 1 \end{pmatrix}, q=1,2 \quad (8)$$

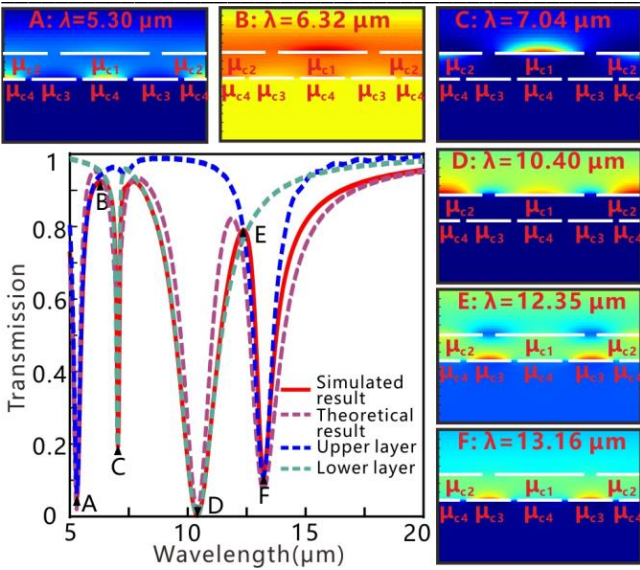

**Figure 3.** Transmission spectra of the proposed GMs based on the FDTD simulation (red solid line) and theoretical model (purple dashed line). The blue dashed line and green dashed line are the transmission spectra of the GMs that only includes the upper GNRs and the lower GNRs, respectively. The normalized magnetic field distributions of the transmission dips (A ($\lambda$=5.30 μm), C ($\lambda$=7.04 μm), D ($\lambda$=10.40 μm) and F ($\lambda$=13.16 μm)) and peaks (B ($\lambda$=6.32 μm) and E ($\lambda$=12.35 μm)).

Under light normally illuminates on the GMs, the Fresnel coefficients in the matrix $M_q$ are expressed as $t_{12}=t_{21}=2n_{SiO2}/(2n_{SiO2}+Z_0\sigma_q')$, $r_{12}=r_{21}=-Z_0\sigma_q'/(2n_{SiO2}+Z_0\sigma_q')$, where $Z_0$=367.7 Ω represents the vacuum impedance and $\varphi' = d_g n_{SiO2}\omega/c$ is the phase difference between the upper GNRs and the lower GNRs. Under the condition of quasistatic approximation, the average sheet conductivity $\sigma_q'$ is given by

$$\begin{cases} \sigma_1' = \dfrac{i}{2}\left( \dfrac{r_1 e^2 \mu_{c1} N\omega}{\pi\hbar^2(\omega^2-\omega_{r1}^2)+i\Gamma_{r1}\omega} + \dfrac{r_1 e^2 \mu_{c2} N\omega}{\pi\hbar^2(\omega^2-\omega_{r2}^2)+i\Gamma_{r2}\omega} \right) \\ \sigma_2' = \dfrac{i}{2}\left( \dfrac{r_2 e^2 \mu_{c3} N\omega}{\pi\hbar^2(\omega^2-\omega_{r3}^2)+i\Gamma_{r3}\omega} + \dfrac{r_2 e^2 \mu_{c4} N\omega}{\pi\hbar^2(\omega^2-\omega_{r4}^2)+i\Gamma_{r4}\omega} \right) \end{cases} \quad (9)$$

where $\omega_{rj}$ is the resonance frequency, which is calculated by using Eq. (6) for different $\mu_{cj}$ (j=1, 2, 3, 4). And the resonance width $\Gamma_{rj}$ of the GNRs is usually 10% larger than the Drude scattering width $\Gamma_j=ev_F^2/(\mu\mu_{cj})$ in the unpatterned graphene, where $v_F\approx c/300$ is Fermi velocity and $\mu$=10000 cm$^2$/V is DC mobility [90]. The phase factor $\Phi_j=m\cdot\varphi_j$ (m=1, 2, 3, 4…) is a fitting parameter deduced from the FDTD simulation. From Eqs. (7)-(9), the transmittance of the GM can be expressed as

$$T=\left[ \dfrac{4n_{SiO2}^2}{(2n_{SiO2}+Z_0\sigma_1')(2n_{SiO2}+Z_0\sigma_2')e^{-i\varphi'}-Z_0^2\sigma_1'\sigma_2'e^{i\varphi'}} \right]^2 \quad (10)$$

According to Eq. (10), the theoretical transmission spectrum of the GMs is shown by the purple dashed line in Figure 3. We find that the theoretical transmission spectrum (purple dashed line) agrees with the simulated transmission spectrum (solid red line) when the fitting parameters $\Phi_j$ are fitted as $\Phi_1$=3.77, $\Phi_2$=0.85, $\Phi_3$=5 and $\Phi_4$=0.45.

We also analyze the influence of structure parameters on the transmission spectrum, and the calculated results are exhibited in Figure 4. As shown in Figure 4(a), the increases of chemical potentials $\mu_{c1}$ and $\mu_{c2}$ lead to the decreases of the resonant wavelengths for dips C and D. Since the generations of dips C and D are mainly originated from the SPPs modes on the GNRs with $\mu_{c1}$ and $\mu_{c2}$, which can be confirmed by normalized magnetic distributions of C and D in Figure 3. And according to Eq. (6), the increase of chemical potential causes the decrease of the effective refractive index, leading to the decrease of resonance wavelength [90]. These two reasons can clearly explain the variation of transmission spectrums in Figure 4(a). Similarly, as shown in Figure 4(b), with the increase of $\mu_{c3}$ and $\mu_{c4}$, the blue-shifts of dip A (D) are induced by the SPPs resonance on the GNRs with $\mu_{c4}$ ($\mu_{c3}$), respectively. As shown in Figure 4(c), when the gap $d_g$ between two layer GNRs increases from 100 nm to 300 nm, dips C and F move to the short-wavelength direction and are more sensitive than dips A and D. As the field distributions of the dips shown in Figure 3, the SPPs resonance of dip C (F) interacts with the vertical graphene ribbons stronger than that of dip A (D), leading to a significant influence of gap $d_g$ on the resonant wavelength of dip C (F) than that of dip A (D). Moreover, as shown in Figure 4(d), the resonant wavelengths of peak E and dips D, F increase significantly when the filling ratio increases from 0.85 to 0.90. The reason for this is as following: as shown in the field distribution in Figure 3, for the graphene ribbons with the same width in the upper GNRs, the resonant SPPs mode for dip D interacts with the horizontal graphene ribbons stronger than that for dip C. As for the graphene ribbons with the same width in the lower GNRs, the SPPs mode for dip F [Re($n_{eff}$)=22.95] is confined more tightly than that for dip A [Re($n_{eff}$)=10.83]. Thus, the resonance wavelengths of dips D and F are more sensitive to the filling ratio than those of dips A and C. According to Eq. (6), the resonance wavelengths of all the dips red-shift with the increase of filling ratio.

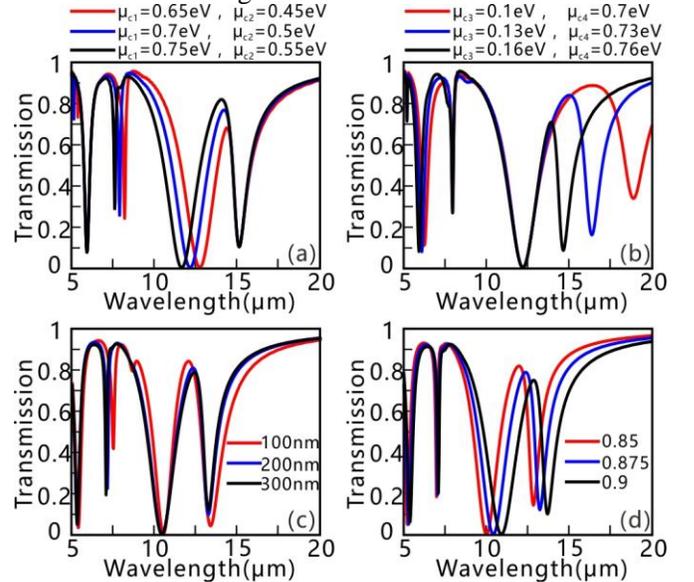

**Figure 4.** The transmission spectra of GMs with the variations of $\mu_{c1}$ ($\mu_{c2}$) (a) and $\mu_{c3}$ ($\mu_{c4}$) (b). $\mu_{c3}$ and $\mu_{c4}$ are set as 0.15 eV and 0.75 eV in (a), respectively. $\mu_{c1}$ and $\mu_{c2}$ are set as 0.7 eV and 0.5 eV in (b), respectively. The influences of gap $d_g$ (c) and filling ratio (d) on the transmission spectrum. Here, $\mu_{c1}$=0.7 eV, $\mu_{c2}$=0.5 eV, $\mu_{c3}$=0.15 eV and $\mu_{c4}$=0.75 eV in (c) and (d). $r_1$=$r_2$=0.9 in (c) and $d_g$ = 300nm in (d).

## 3. Spectrum prediction and inverse design

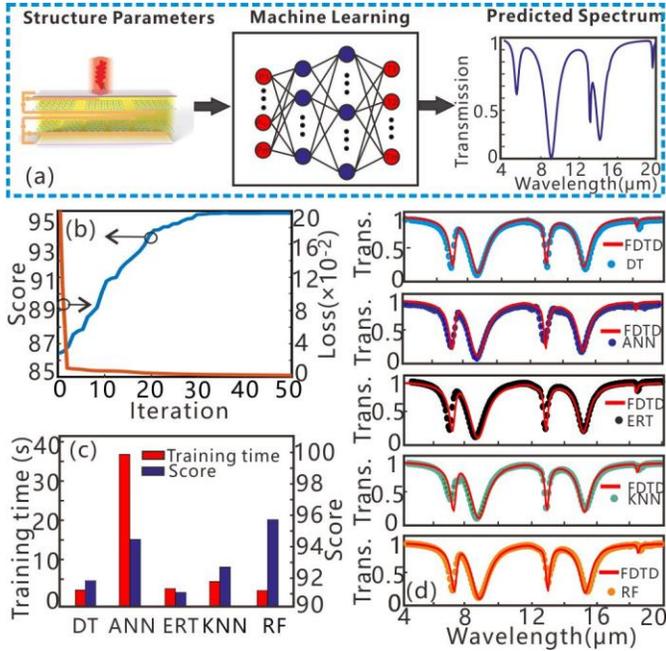

**Figure 5.** (a) The diagram of the forward spectrum prediction. (b) Score and loss for different generations of the GA. (c) Training time and accuracies for different regression algorithms in forward spectrum prediction. (d) The transmission spectrums predicted by the regression algorithms and simulated by the FDTD simulation.

As shown in Figure 4, it can be found that the slight changes of the structure parameters have significant influence on the transmission spectrum. If we want to discover the potential relationship between the structure parameters of the GMs and the transmission spectrum, it requires a high computational cost to traverse all structure parameters. Although we have employed 2D FDTD simulation to reduce simulation time, it requires several minutes to guarantee the convergence of simulation. Actually, we can use Monte Carlo simulation or interval sampling method to reduce simulation time, but it leads to the loss of accuracy due to interpolation and fitting. Another way to improve the efficiency is to train a model based on machine learning algorithms by using a small part of simulation results [28-50]. It had been proven that ANNs-based models could equivalently replace the electromagnetic simulation for some photonic structures and the inference time of ANNs is far less than that of the simulation methods [28-50]. The prediction process for transmission spectrum according to the structure parameters is known as 'forward spectrum prediction' [20]. By contrast, inverse design based on the ANNs trains a reliable regression model to search for the most suitable structure parameter for a given transmission spectrum. The ANNs used in the inverse design have been demonstrated to compete with the traditional optimization algorithms [20]. Notably, the principles behind the forward spectrum prediction and inverse design based on the ANNs are date regression between the structure parameters and electromagnetic response. It means that the labels of training data are continuous variables rather than discrete variables. There are several machine learning algorithms can be used in data regression. Compared with the classical machine learning algorithms, such as SVM and RF, the training cost of the ANNs is much higher. It has been demonstrated that SVM performs better than the ANNs in the trend prediction of soil organic carbon and river flow [60-61]. In addition, the selection of the hyper-parameters for ANNs (layers, solver, activation function, learning rate, batch size and so on) is more complex than that of SVM and RF. The evolution algorithms are used to find the optimal hyper-parameters for the ANNs [20]. In addition, the training time and inference time of the ANNs significantly exceed those of the classical machine learning algorithms [60]. The reason for this is related to the training process of the ANNs includes forward-propagation, back-propagation and stochastic gradient decent, while the training processes of classical machine learning algorithms are relative simple [61]. In order to overcome the defects of the ANNs, we use several classical regression algorithms to achieve the forward spectrum prediction and inverse design for the GMs. Similar to kNN classification, kNN regression calculates the distances between the targeted instance and each training instance and then selects the most similar $k$ data as candidate set to determine the results [91]. And three tree-based regression algorithms, including DT, RF, and ERT, are also used in the inverse design for GMs. These tree-based regression algorithms have the same steps, such as selecting splits and selecting the optimal tree [92]. RF is an ensemble algorithm based on the bootstrap aggregating that involves several regression trees [93]. Compared with the RF, the split of features for the ERT is more random, leading to the reduction of the variance for the trained model [94].

First of all, we attempt to use the regression algorithms to replace the FDTD simulation in the forward spectrum prediction. In Figure 5(a), it can be found that the regression algorithms take the structure parameters of the GMs as the input and predict corresponding transmission spectrum. For instance, the potential relationships between the chemical potentials of graphene ribbons $\mu_{c1}$, $\mu_{c2}$, $\mu_{c3}$ and $\mu_{c4}$ and the transmittances in the transmission spectrum are taken into consideration. In order to train the regression models, we use the repeatable FDTD simulation and Monte Carlo simulation to generate training sets because the regression algorithms belong to supervised learning [95]. Each instance in 20,000 training instances includes 4 structure parameters ($\mu_{c1}$, $\mu_{c2}$, $\mu_{c3}$, $\mu_{c4}$) and 200 transmittances evenly sampled from the transmission spectrum. And all structure parameters are initialized in different ranges specified by minimum and maximum values 0.6 eV<$\mu_{c1}$<0.8 eV, 0.4 eV <$\mu_{c2}$< 0.6 eV, 0.05 eV<$\mu_{c3}$< 0.25 eV and 0.6 eV<$\mu_{c1}$<0.8 eV. It means that the chemical potentials of graphene ribbons are randomly generated from the ranges with the precision of 0.1 eV. When we have enough training instances, the models based on regression algorithms are trained by using 20000 training instances, while another 2000 instances are left as the test set to validate the training effect. It should be noticed that although the generation of 22000 training instances takes 23 hours by using a high performance server, the prediction speed of regression algorithms for a new structure parameter is faster than the 2D FDTD simulation once the model are constructed [28]. Before training the model, we should pay attention to the influence of hyper-parameters on the training effect. For instance, we should consider the number of trees in the forest and the maximum depth of tree for the RF. Here, the deterministic process of hyper-parameters for the ANNs is relatively complex because there are a great deal of hyper-parameters should be considered [20]. In order to achieve a good accuracy, we use GA to find optimal hyper-parameters and network architecture for the ANNs, and the variations of the loss and accuracy for different generations are shown in Figure 5(b). Here, the accuracies of the regression algorithms are represented by the scores which measure the similarity between the predicted transmission spectrums and practical transmission spectrums (the best and worst values for the score are 1 and arbitrary negative, respectively) [96]. And the scores are regarded as the fitness for the GA used in finding the optimal hyper-parameters for ANNs. As shown in Figure 5(b), the score (loss) increases (decreases) from 86.8 (20) to 95 (0.01), meaning that the optimization of hyper-parameters for the ANNs are efficient. In addition, we also use the same training set to train other regression algorithms.

Figure 5(c) shows the training time and accuracies for different regression algorithms. Surprisingly, it can be found that the scores of all regression algorithms are greater than 91, indicating that other regression algorithms are competitive with the ANNs in the forward spectrum prediction. Although the ANNs-based model is effective intuitively, the accuracy (score) of RF (96) outperforms that of ANNs (95). In order to illustrate the effectiveness of regression models vividly, we compare the transmission spectrums predicted by regression algorithms and simulated by the 2D FDTD simulation. We randomly select a group of structure parameter from the test set and calculate transmission spectrums based on regression algorithms and numerical simulation. As shown in Figure 5(d), the transmission spectrums predicted by the regression algorithms agree with the FDTD simulation results obviously. Compared with the transmission spectrum predicted by the ANNs, the transmission spectrum predicted by RF are closer to the ground truth (the transmission spectrum calculated by the FDTD simulation). More importantly, once the hyper-parameters are determined, the training cost of the ANNs (36 seconds) exceeds those of other machine learning algorithms, especially for the RF. With a comprehensive consideration of training cost and accuracy, RF regression algorithm is a more appropriate method to complete forward spectrum prediction for the GMs in comparison to the ANNs.

Similar to forward spectrum prediction, the regression algorithms mentioned above can be employed in the inverse design for the GMs. Contrary to forward spectrum prediction, Figure 6(a) shows the diagram of the inverse design for the GMs. It can be found that the inputs and outputs of the models based on regression algorithms are the transmittances in transmission spectrum and the structure parameters of the GMs, respectively. It should be noted that there is no need to generate new training instances, we use the same training instances to train the model by reversedly converting the inputs (outputs) to outputs (inputs) in the forward spectrum prediction. The training time and accuracies for all regression algorithms in the inverse design are exhibited in Figure 6(b). It can be found that all the regression algorithms can achieve excellent performance and the score of DT (93) is lower than that of ANNs (97), ERT (96), kNN (96.5) and RF (98). To validate the effectiveness of the regression algorithms in the inverse design, we randomly select a transmission spectrum from the test set and input it into the model. The structure parameters (chemical potentials $\mu_{c1}$, $\mu_{c2}$, $\mu_{c3}$ and $\mu_{c4}$ of the GNRs) of the GMs predicted by regression algorithms and the ground truth are shown in Figure 6(c). We can observe that the predicted chemical potentials $\mu_{c1}$, $\mu_{c2}$, $\mu_{c3}$ and $\mu_{c4}$ are close to the practical chemical potentials (red dashed line), confirming the effectiveness of all the regression algorithms. In addition, we also use the chemical potentials predicted by the regression algorithms to simulate the GMs based on the FDTD simulation. As shown in Figure 6(d), the accuracy of RF outperforms that of DT, ANNs, ERT and kNN because high similarity. More importantly, the training time of the RF (6 seconds) is lower than that of the ANNs (34 seconds). Obviously, the calculated results shown in Figure 6(b)-(d) indicate that the ANNs is not the best choice for inverse design of the GMs. And the RF outperforms the ANNs in terms of accuracy and efficiency. It should be noted that although the performance of the RF is superior to that of the ANNs for the inverse design of the GMs. That does not mean the RF is a better choice for the inverse design of photonic device. The choice of machine learning algorithms usually depends on the application scenarios. For the inverse design of photonic devices, the ANNs, especially for deep learning, may be more effective for the complicated photonic devices. But for the uncomplicated application scenarios, such as the inverse design and optimization for a photonics device that contains a small number of structure parameters (<15), the ANNs may not be the best choice.

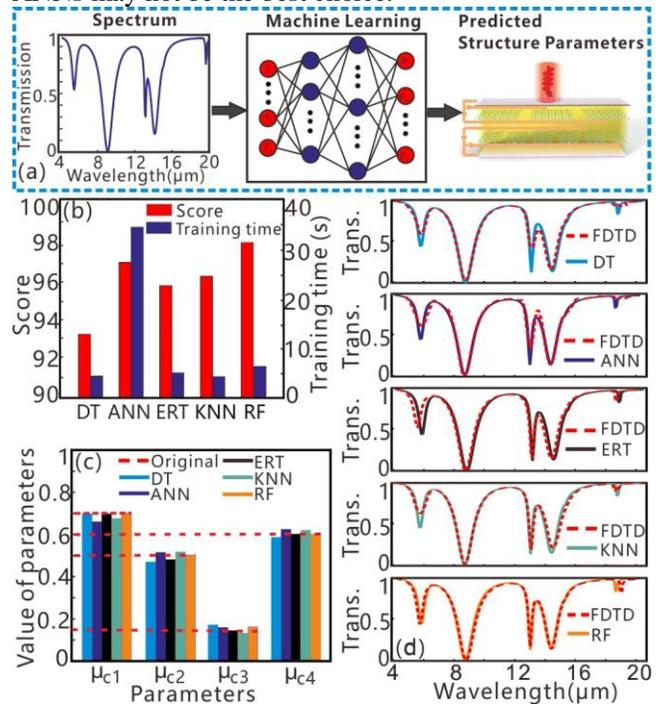

**Figure 6.** (a) The diagram of the inverse design. (b) The training time and accuracies (scores) for all regression algorithms in inverse design. (c) The structure parameters (chemical potentials $\mu_{c1}$, $\mu_{c2}$, $\mu_{c3}$ and $\mu_{c4}$ for the GNRs in the GMs) predicted by all regression algorithms and the ground truth. (d) The FDTD simulated transmission spectrums for the chemical potentials predicted by the regression algorithms.

## 4. Performance optimization

The performance optimization of the transmission spectrum are usually divided into two categories: optimization for the wind bandwidth transmission spectrum and optimization for the performance metrics, such as transmittance or bandwidth. As same as the inverse design, the optimized transmission spectrum with a wide wavelength range is comprehensively optimized by inputting it into the data-driven model. Similar to inverse design, the data-driven model predicts the physical parameters, which can generate the optimized transmission spectrum. Notably, if the optimized transmission spectrum is not achievable, the data-driven model may predict an absurd physical parameter. On the other hand, the performance metrics of transmission spectrum, such as the transmittance at a given wavelength or the bandwidth of a transparency window can be pertinently optimized. It's difficult to solve the optimization problem of performance metrics by using the data-driven model. While the gradient free methods, such as GA and particle swarm optimization (PSO), have been applied in the optimization of performance metric because of their simplicity and effectiveness [67-71]. In this section, we try to use the evolution algorithms to optimize the GMs and compare the optimization results between different evolution algorithms, including the single-objective optimization and the multi-objective optimization.

The algorithmic details of the GA are outlined as follows: (i) randomly generating an initial population consisted of $N$=40 individuals. Each individual has four structure parameters, namely, the chemical potentials of graphene ribbons ($\mu_{c1}$, $\mu_{c2}$, $\mu_{c3}$, $\mu_{c4}$). Here, all structure parameters are initialized in different ranges specified by minimum and maximum values 0.6 eV<$\mu_{c1}$<0.8 eV, 0.4 eV<$\mu_{c2}$< 0.6 eV, 0.05 eV<$\mu_{c3}$< 0.25 eV and 0.6 eV<$\mu_{c4}$<0.8 eV. (ii) For the $N$ group of structure parameters, transmission spectrums are simulated by using the FDTD method. And different performance metrics, such as the transmittance at a given wavelength, are regarded as the optimization objective and fitness for the GA. If the transmission spectrum with a wide bandwidth is optimized, the fitness can be defined as

$$F=\sum_{\lambda_{\min}}^{\lambda_{\max}}\left|S_0(\lambda)-S(\lambda)\right| \quad (11)$$

where $\lambda$, $\lambda_{min}$ ($\lambda_{max}$), and $S_0(\lambda)$ ($S(\lambda)$) are the wavelength, minimum (maximum) wavelength and targeted (optimized) transmission spectrum, respectively. Then, the individuals of population are sorted according to the fitness in descending order. (iii) Trying to generate a new population by using standard selection, crossover and mutation procedures. In the selection process, two parent individuals are selected from the previous generation based on roulette-wheel selection method or tournament strategy [97]. The structure parameters with better fitness are selected with higher probability. To maintain the diversity of population and keep some superior individuals, some percentage of the superior (inferior) structure parameters are kept in the next generation. In the crossover process, structure parameters are converted into binary values firstly. It should be noted that the conversion of decimal to binary is likely to result in the loss of digital precision. After that, the optimization variables (structure parameters) of parent individuals cross over to generate a new population based on uniform crossover algorithm or single-point crossover [97]. In the mutation process, each element in binary number has 5% probability to flip from 0 (1) to 1 (0). After converting the optimization variables (structure parameters) from binary number to decimal number, a new population is generated. (iv) The fitness of newly generated population are evaluated to determine the optimization process whether stop or not. If the generation of structure parameters evolve for 1000 times or optimization objective remain unchanged for more than 5 generations, the GA stops, otherwise, proceeds to Step (ii). Quantum genetic algorithm (QGA) is a new parallel evolution algorithm, which combines with traditional GA and quantum algorithms [98]. In the QGA, the encode method for the variable is quantum bit rather than binary number. And in the crossover and mutation processes, QGA uses the quantum rotation gate to update the individual.

Similar to the GA, PSO is an evolution algorithm which is suitable for decimal number rather than binary number [69]. The generation of initial population for PSO is the same as that of the GA. However, the generation of new population for PSO is not through selection, crossover and mutation procedures. It means that there no need to convert the decimal number to binary number in the PSO, which can effectively avoid the loss of digital precision. For PSO, the individuals in the population depend on the globally optimal individual and historically optimal record for each individual to search for the optimal solution [69]. Similarly, when we use PSO to optimize the GM, each individual in population searches for the optimal structure parameters by synthetically considering currently optimal structure parameters (OPS) and individually OPS. The evolution of the structure parameters is controlled by a specified velocity [69]:

$$V_i^{k+1} = WV_i^k + c_1 r_1 \left(pb_i^k - X_i^k\right) + c_2 r_2 \left(gb_k^d - X_i^k\right) \quad (12)$$

where $i$ is the $i$th structure parameter in the population, $k$ is the iteration number, $W$ is the inertia weight, $c_1$=$c_2$=1.49445 are the acceleration constants, $r_1$ ($r_2$) is the random value between 0 and 1, $gb_k^d$ is the globally OPS, $X_{ik}$ and $pb_i^k$ are current structure parameter and individually OPS for the $i$th structure parameter in the $k$th iteration, respectively. The $i$th structure parameter in the population is updated according to following equation:

$$X_i^{k+1} = X_i^k + V_i^{k+1} \quad (13)$$

In order to avoid the premature problem, the velocities of evolution are limited to a certain range (-1~1). Finally, the evaluation of the newly generated population is the same as that of the GA. If the population does not meet the termination conditions, the velocities of structure parameters are calculated based on Eq. (12) in the next iteration. To compare the optimization effects between the GA, QGA and PSO, we randomly select a complete transmission spectrum (red dashed line in Figure 7(c)) in test set as optimization objective. For all optimization algorithms, the degree of approximation (Eq. (11)) between targeted transmission spectrum and optimized transmission spectrum is treated as the fitness for the GA, QGA and PSO. Figure 7(a) shows the fitnesses of the GA, QGA and PSO for different generations in the optimization of the targeted transmission spectrum. It can be observed that the fitnesses of the GA, QGA and PSO are close to 0, indicating these single-objective optimization algorithms are convergence. And the convergence speeds of the GA and PSO are faster than that of the QGA. Here, in the 100th generation, we select the optimized chemical potentials for all optimization algorithms and compare them with the ground truth. It can be found in Figure 7(b) that the chemical potentials optimized by the GA, QGA and PSO agree well with the targeted chemical potentials. In Figure 7(c), the green solid line and blue solid line are the optimized transmission spectrums in the first generation and in the 100th generation, respectively. It can be observed that the optimized transmission spectrums in the first generation (green solid lines) are randomly generated and those in the 100th generation (blue solid lines) are close to the targeted transmission spectrums.

Finally, the GMs is optimized for several performance metrics, such as the transmittances at different wavelengths. The steep degree of the PIT effect is a critical performance indicator, which affects the bandwidth, group index, figure of merit and so on. To achieve steep transmission characteristics, we use a multi-objective optimization algorithm, namely non-dominated sorting genetic algorithm-II (NSGA-II), to optimize for the transmittances at different wavelengths. Compared with other multi-objective optimization algorithm, NSGA-II finds the pareto optimal solution based on the fast nondominated sorting method (FNSM) and elitist strategy [99]. In the NSGA-II, the crowding distances of individuals and the levels calculated by the FNSM are combined to jointly determine the order of the individuals [99]. For all performance indicators, the individuals in the lower level are better than those in the higher level, while the individuals in the same level are incommensurable. In our simulations, the steep degree of the PIT effect is simply characterized as the differences between the transmittances of transmission peaks and dips. The algorithmic details of the NSGA-II are outlined as follows: (i) the generation of initial population for the NSGA-II is the same as that of the GA, QGA and PSO. Here, each individual has seven structure parameters, namely, the chemical potentials of graphene ribbons ($\mu c_1$, $\mu c_2$, $\mu c_3$, $\mu c_4$), the filling ratio of graphene ribbons ($r_1$, $r_2$) and the distance $d_g$ between the upper GNRs and the lower GNRs. Here, all structure parameters are initialized in different ranges 0.6 eV$<\mu c_1<$0.8 eV, 0.4 eV$<\mu c_2<$ 0.6 eV, 0.05 eV$<\mu c_3<$ 0.25 eV, 0.6 eV$<\mu c_4<$0.8 eV, 0.7$<r_1<$0.9, 0.7$<r_2<$0.9 and 100 nm$<d_g$ $<$300 nm. (ii) The differences between the transmittances at different wavelengths are regarded as the fitness for the NSGA-II. It means that two differences and four differences between the transmission peaks and dips are calculated for one transparency window and two transparency windows, respectively. Unlike the GA, QGA and PSO, the levels of the individuals in the population for the NSGA-II are determined by the FNSM. And the crowding distances are calculated for the individuals in the same level to maintain the diversity of the population. The individuals in the population are sorted according to the levels and crowding distances [99]. (iii) The generation process of a new population for the NSGA-II is the same as that of the GA, QGA and PSO. (iv) The individuals in the newly generated population are placed into the old population to generate a large population. And the individuals in the large population are sorted based on the FNSM and crowding distances. Finally, top $N$ individuals are selected to generate a new population for next iteration based on elitist strategy [99]. (v) The evaluation of newly generated population for the NSGA-II is similar to the GA and PSO. And the best individual in the pareto front is selected as the solution for the NSGA-II. Figure 7(d) and Figure 7(e) exhibit the multi-objective optimization results for one transparency window and two transparency windows, respectively. Here, the optimization objective for one transparency window is two differences between a transmission peak and two dips, while that for two transparency windows is four differences between two transmission peak and four dips. After 100 iterations, the differences between the transmission peaks and dips reach to 0.76 and 0.97 (0.87, 0.83, 0.79 and 0.69) for one transparency window (two transparency windows), indicating the NSGA-II is effective for the optimization of the GM. Obviously, the multi-objective optimization can be used to achieve steep transmission characteristics by synthetically considering several performance metrics.

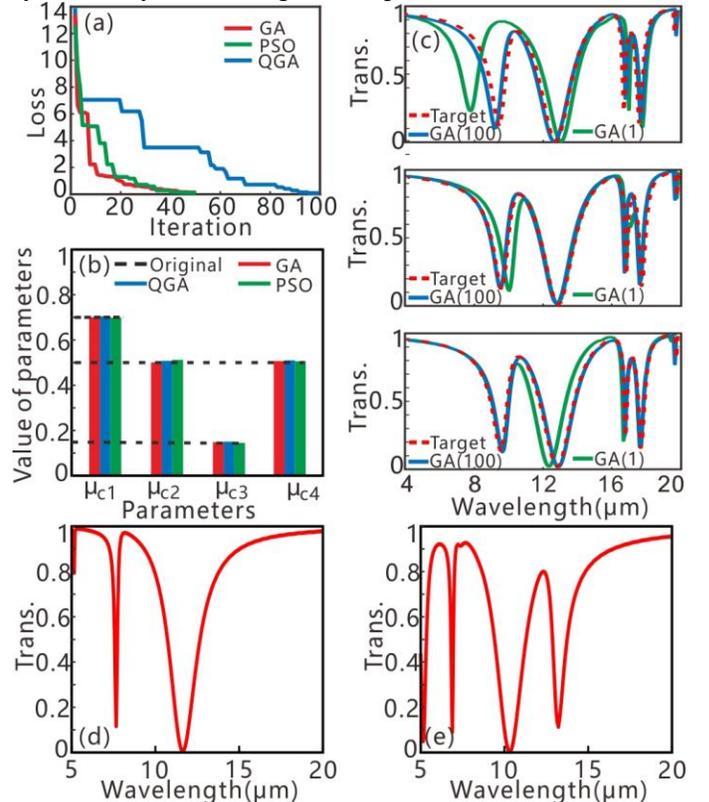

**Figure 7** (a) The fitnesses of the GA, QGA and PSO for different generations in the performance optimization. (b) Optimization results of chemical potentials for the GA, QGA and PSO in the 100th iteration. (c) The optimized transmission spectrums of the GA, QGA and PSO in the first iteration (green line) and the 100th iteration (blue line). (d) The multi-objective optimization results for two differences between one peak (8161 nm) and two dips (7659 nm and 11620 nm). (e) The multi-objective optimization results for four differences between two peaks (6110 nm and 12620 nm) and four dips (5150 nm, 6890 nm, 10310 nm and 13220 nm).

## 5. Conclusion

In this article, we propose an intelligent approach to achieve forward spectrum prediction, inverse design and performance optimization for the GMs. The structure parameters of the GMs are well-designed to obtain PIT effect in transmission spectrum. The theoretical transmission spectrum based on transfer matrix method agree well with the FDTD simulated transmission spectrum. In addition, several classical machine learning algorithms are used to achieve the forward spectrum prediction and inverse design for the GMs. The calculated results demonstrate that all the algorithms are effective and the RF has advantages in terms of accuracy and training speed in comparison to the ANNs. Moreover, we use the single-objective optimization algorithms and multi-objective optimization algorithms to optimize for the GMs by taking many performance metrics into consideration synthetically. The maximum difference between the transmission peaks and dips in optimized transmission spectrum can reach 0.97. This work paves a new way towards the intelligent design for graphene-based structures and has important applications in the design of advanced materials and metamaterials.

## Acknowledgments

National Natural Science Foundation of China (61705015, 61625104,61431003); Beijing Municipal Science and Technology Commission (Z181100008918011); Fundamental Research Funds for the Central Universities (2019RC15, 2018XKJC02); National Key Research and Development program (2016YFA0301300).